\documentclass[aps,prl,amsmath,amssymb,reprint,twocolumn]{revtex4-2}
\usepackage{amsfonts,amsthm,amsmath,amssymb,graphicx,color}
\usepackage{mathrsfs}
\usepackage{hyperref}
\usepackage{float}
\usepackage{subcaption}
\usepackage{hhline}
\usepackage{bbold}

\begin{document}
\title{A measure of chaos from eigenstate thermalization hypothesis}
\author{Nilakash Sorokhaibam}
\email{ns15485863@gmail.com}
\affiliation{Department of Physics, Tezpur University, Tezpur, 784028, Assam, India.}

\begin{abstract}
Eigenstate thermalization hypothesis is a detailed statement of the matrix elements of few-body operators in energy eigenbasis of a chaotic Hamiltonian. Part of the statement is that the off-diagonal elements fall exponential for large energy difference. We propose that the exponent ($\gamma>0$) is a measure of quantum chaos. Smaller $\gamma$ implies more chaotic dynamics. The chaos bound is given by $\gamma=\beta/4$ where $\beta$ is the inverse temperature. We give analytical argument in support of this proposal. The slower exponential fall also means that the action of the operator on a state leads to higher delocalization in energy eigenbasis. Numerically we compare two chaotic Hamiltonians  - SYK model and chaotic XXZ spin chain. Using the new measure, we find that the SYK model becomes maximally chaotic at low temperature which has been shown rigorously in previous works. The new measure is more readily accessible compare to other measures using numerical methods.
\end{abstract}

\maketitle

Tremendous advances have been made in identifying new diagnostics of quantum chaos and measures of quantum chaos in the last two decades. Traditionally, quantum chaos is diagnosed using level statistics \cite{Bohigas:1983er}. In recent years, many other diagnostics have also been proposed. Some of the well known ones are exponential decay of out-of-time-ordered (OTO) correlators \cite{Maldacena:2015waa}, spectral form factor \cite{Papadodimas_2015,Cotler_2017} and many others \cite{Kudler_Flam_2020}. Diagnostic like OTO correlators has been shown to be superior to the study of level statistics in certain systems \cite{Akutagawa_2020}. Some of the diagnostics can even quantity the strength of chaotic dynamics. The most popular one is the Lyapunov exponent from OTO correlators. But it has also been shown that scrambling dynamics which is responsible for the exponential decay of OTO correlators does not always come from quantum chaos \cite{Xu_2020,Dowling_2023}. All-in all, study of level statistics seems to be the most conservative diagnostic of quantum chaos. On the other hand, thermalization in quantum systems is guaranteed by eigenstate thermalization hypothesis (ETH) \cite{PhysRevA.43.2046,Srednicki:1994mfb}. ETH has been shown to hold true for many well-known quantum chaotic systems. Whether ETH and quantum chaos implies each other is still not completely settled but the available evidences overwhelmingly suggest that ETH and quantum chaos implies each other. For the rest paper, we will be working with the assumption that ETH and quantum chaos implies each other.

In this paper, we will identify a measure of chaos from ETH. The Lyapunov exponent $\lambda$ has a bound $\lambda\leq 2\pi/\beta$ where $\beta$ is the inverse temperature \cite{Maldacena:2015waa}. This bound has also been derived from ETH \cite{Murthy:2019fgs}. ETH statement is concerned with local operators which correspond to physical observables. The precise statement of ETH is that the matrix elements of such operator $A$ in energy eigenbasis $|E_i\rangle$ is given by
\begin{equation}
A_{ij}=\mathcal{A}(E)\delta_{ij}+e^{-S(E)/2} f(E,\omega)R_{ij}\
\label{ETH}
\end{equation}
where $|i\rangle$ and $|j\rangle$ are energy eigenstates with energies $E_i$ and $E_j$ respectively and $E=(E_i+E_j)/2$, $\omega=E_i-E_j$. $S(E)$ is the entropy at energy $E$. $\mathcal{A}(E)$ is equal to the expectation value of $A$ in the microcanonical ensemble at energy $E$ or other ensembles by the equivalence of ensembles. $f(E,\omega)$ is a smooth function of its arguments. $f(E,\omega)$ also falls exponentially for large $|\omega|$ \cite{Murthy:2019fgs}. $f(E,\omega)$ is also a monotonically increasing function of $S(E)$ \cite{sorokhaibam2024quantum}. $R_{mn}$ are pseudo-random variables with zero mean and unit variance. $A$ is hermitian and hence $R_{ij}=R^*_{ji}, \; f(E,-\omega)=f^*(E,\omega)$. The statistical average of $R_{ij}R_{jk}R_{kl}R_{li}$ are given by \cite{Foini:2018sdb}
\begin{equation}
\overline{R_{ij}R_{jk}R_{kl}R_{li}}=\delta_{ik}+\delta_{jl}+e^{-S(E)}g(E,\omega_1,\omega_2,\omega_3)
\end{equation}
where $E=(E_i+E_j+E_k+E_l)/4$, $\omega_1=E_i-E_j$, $\omega_2=E_j-E_k$, $\omega_3=E_k-E_l$. Note that the extra $e^{-S(E)}$ in $g$ term is in addition to the $e^{-S(E)/2}$ suppression in (\ref{ETH}). For simplicity, we have used an operator $A$ with thermal expectation value equal to zero $\mathcal{A}(E)=0$.

Our proposal is concerned with the rate of exponential fall of $f(E,\omega)$ for large $|\omega|$. We propose that

\begin{center}
\emph{the rate of exponential fall of $f(E,\omega)$ for large $|\omega|$ is a measure of quantum chaos, with a lower bound given by $\beta/4$ which is the maximal chaos limit.}
\end{center}

A slower exponential fall corresponds to more chaotic dynamics. Mathematically,
\begin{equation}
f(E,\omega) \sim e^{-\gamma|\omega|}, \quad \gamma\geq \beta/4, \qquad \text{for large } |\omega|\
\label{result}
\end{equation}

Now we will motivate why $\gamma$ can be used as a measure of chaos. We will suppress $E$ dependence from here on. For this, we will closely follow the derivation of Lyapunov bound from ETH in \cite{Murthy:2019fgs}. Consider the system is in a thermal state with density matrix given by $\rho=e^{-\beta H}$ where $H$ is the Hamiltonian of the system. Finiteness of $A^2$ expectation value in large system-size limit enforces that $f(\omega)$ should fall exponentially for large $|\omega|$ at least as fast as
\begin{equation}
f(\omega) \sim e^{-\beta|\omega|/4}\
\label{fol}
\end{equation}
Note that random matrix theory (RMT) violates this lower bound on the exponential fall. In RMT, $f(\omega)$ is a constant.

The rate of exponential fall of $g(\omega_1,\omega_2,\omega_3)$ is fixed by the connected part of the four point function
\begin{gather}
F_4(t_1,t_2,t_3)=\text{Tr} [\rho^{1/4}A(t_1)\rho^{1/4}A(t_2)\rho^{1/4}A(t_3)\rho^{1/4}A(0)]\
\end{gather}
and, assuming (\ref{fol}), it has been found to be
\begin{equation}
g(\omega_1,\omega_2,\omega_3)\sim e^{-\beta|\omega_3|/4} \quad \text{for fixed  } \omega_1, \omega_2\
\label{gol}
\end{equation}
The OTO correlator from which the Lyapunov exponent can be extract is the connected part of $F_4(t,0,t)$ which is denoted by $F_{4C}(t)$. The Fourier transform of $F_{4C}(t)$ is given by
\begin{eqnarray}
\tilde{F}_{4C}(\omega)&=&\int_{\omega_1,\omega_2}f(\omega_1)f(\omega_2)f(\omega-\omega_1)f(-\omega-\omega_2)\nonumber\\
&&\qquad\qquad \times g(E,\omega_1,\omega_2,\omega-\omega_1)\
\end{eqnarray}
The chaos bound ($\lambda\leq 2\pi/\beta$) is found from the exponential fall of $\tilde{F}_{4C}(\omega)$ for large $|\omega|$.

Now if $f(E,\omega)$ falls faster than what is specified in (\ref{fol}), meaning in (\ref{result}) $\gamma>\beta/4$, then $g(\omega_1,\omega_2,\omega_3)$ perhaps can fall slower than what has been set as a slowest possible limit in (\ref{gol}). But note that what has been set in (\ref{gol}) is the slowest exponential fall, so it is most likely that even for a faster exponential fall of $f(E,\omega)$, the slowest possible fall of $g(E,\omega_1,\omega_2,\omega_3)$ is still fixed by $(\ref{gol})$. In other words, we take the bound on the exponent in (\ref{gol}) to be universal.

Going through the standard steps in \cite{Murthy:2019fgs}, we find that the relation between $\gamma$ and $\lambda$ is
\begin{equation}
\gamma = \frac{3\pi}{4\lambda}-\frac{\beta}{8}\
\end{equation}
This is the relation between our new measure and the well-known Lyapunov exponent $\lambda$.

In the rest of the paper, we will show that the highly chaotic SYK model satisfies our new bound in the low temperature limit. We will also consider a chaotic XXZ spin chain and show that they are less chaotic compare to SYK model. The advantage of our new measure is that it is ready accessible numerically. On the other hand, calculation of Lyapunov exponent from OTO correlators becomes murky when timescales of scrambling and dissipation are not well separated.

The Sachdev-Ye-Kitaev (SYK) model \cite{Sachdev:1993,Kitaev:2015,Maldacena:2016hyu,Chowdhury_2022} has been intensely studied for the last decade or so. The model has many remarkable properties. One of its important properties is that it saturates the chaos bound at low temperature. The SYK model is a quantum system of a large $L$ number of fermions with random all-to-all interaction. The interaction can be between any even q number of fermions. So, it is a non-local or a 0-dimensional system. We will work with 4-fermion (q=4) interaction. The observable that we will be considering is the occupation number operator of a site (say site-1). The Hamiltonian of SYK model and the operator $A=n_1$ are
\begin{gather}
H_{SYK}=\sum_{i,j,k,l=1}^{L}J_{ijkl}\Psi_i^{\dagger}\Psi_j^{\dagger}\Psi_k\Psi_l, \quad A_{SYK}=n_1=\Psi_1^{\dagger}\Psi_1\
\label{HSYK}
\end{gather}
The couplings $J_{ijkl}$ are random numbers drawn from a Gaussian distribution with zero mean and variance equal to $6/L^3$. In many numerical works on SYK model, the quantities of interest are averaged over large of realizations of the couplings. But here, we will be considering only one realization so no averaging will be performed.

The XXZ spin chain is chaotic with a large next-to-nearest neighbour interaction \cite{Gubin_2012}. We consider system of size $L$. The observable that we will be considering is the spin of a site (say site-1), $S^z_1$. We choose open boundary condition to break translational invariance. The Hamiltonian of XXZ spin chain is
\begin{gather}
H_0=\sum_{i=1}^{L-1}\left[J_{xy}\left(S_i^x S_{i+1}^x+S_i^y S_{i+1}^y\right)+J_zS_i^z S_{i+1}^z\right]\nonumber\\
+\sum_{i=1}^{L-2}J'_zS_i^z S_{i+2}^z, \qquad A_{XXZ}=S^z_1\
\end{gather}

\begin{figure}
\centering
\includegraphics[width=1.\columnwidth]{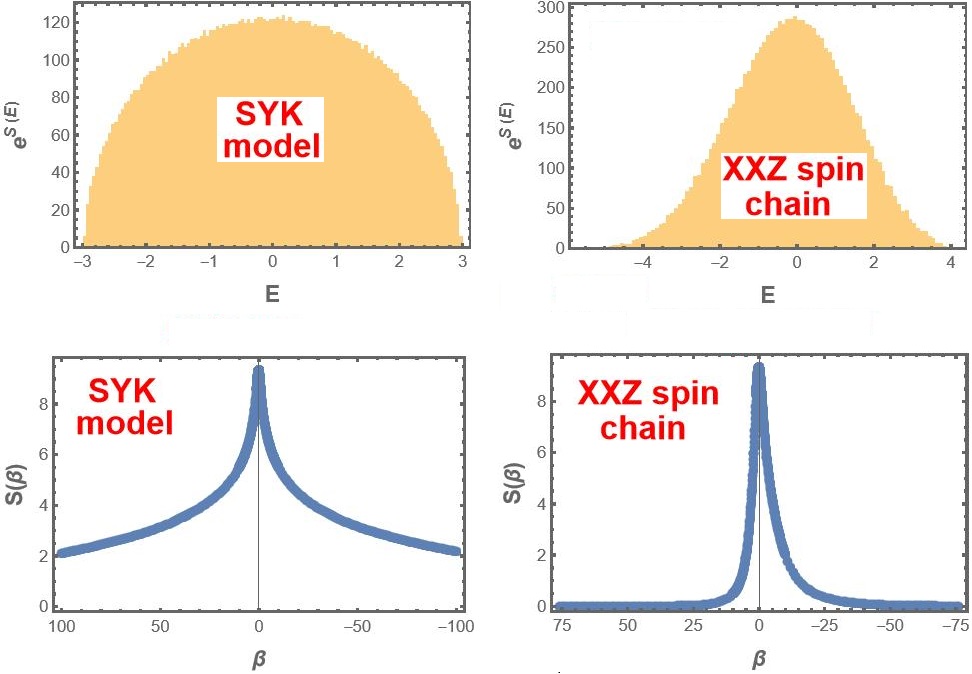}
\caption{\small{The top panels are plots of the spectrum of the two models. The lower panels are plots of the entropy as a function of the inverse temperature $\beta$.}}
\label{EnSplot}
\end{figure}

The top panels in Figure \ref{EnSplot} are plots of the spectrum of the two models. The lower panels are plots of the entropy as a function of the effective inverse temperature $\beta_{eff}$. As, one can see the entropy of SYK model is very large even for very low temperature. This is the manifestation of the well-known result that the entropy of SYK model is non-zero at zero temperature (when taking the limit $L\to\infty$ first and then taking $\beta\to 0$).

\begin{figure}
\centering
\includegraphics[width=1.\columnwidth]{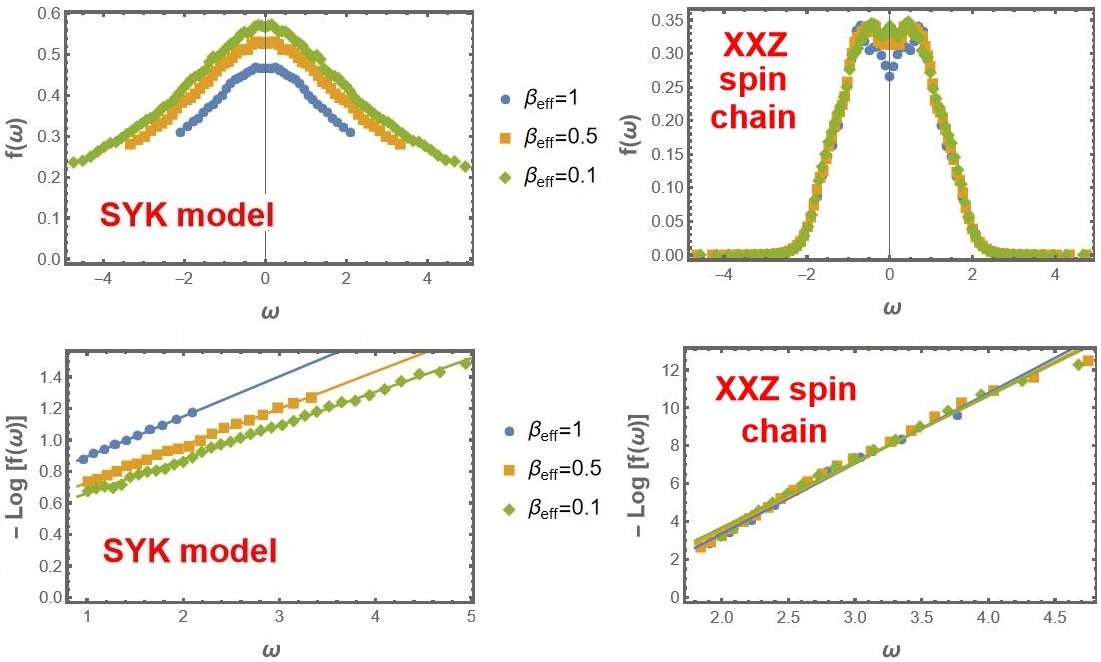}
\caption{\small{The top panels are plots of $f(\omega)$ as a function of $\omega$ for fixed $E$ which corresponds to effective inverse temperature $\beta_{eff}$. The lower panels are plots of $-\log[f(\omega)]$ as a function of large $\omega$ for the same data set.}}
\label{fo_o_plot}
\end{figure}

Next we calculate the $f(\omega)$ as a function of $\omega$ for fixed $E$ which corresponds to effective inverse temperature $\beta_{eff}$. The top panels of Figure \ref{fo_o_plot} are the plots. The lower panels are plots of $-\log[f(\omega)]$ as a function of large $\omega$. They are nice straight lines. In the small temperature limit, we will find that the slope is given by $\gamma=\beta/4$ for SYK model. In case of XXZ spin chain, the exponent does not change much with the variation of temperature. Note that it is evident from the top panels that $f(E,\omega)$ is bigger for higher positive temperature which means that $f(E,\omega)$ is a monotonically increasing function of $S(E)$ \cite{sorokhaibam2024quantum}.

\begin{figure}
\centering
\includegraphics[width=1.\columnwidth]{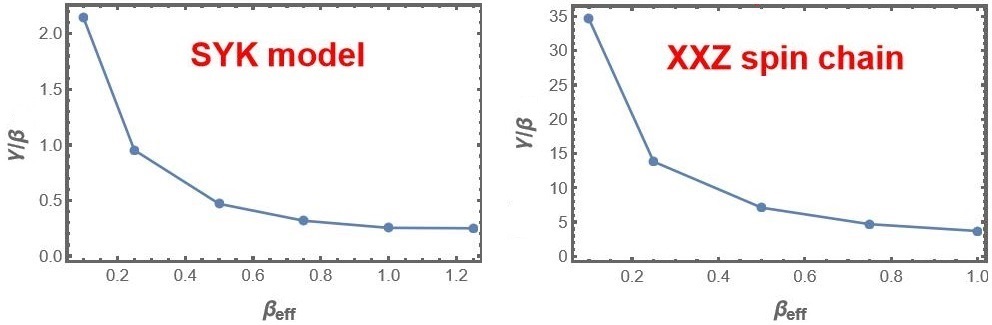}
\caption{\small{Plot of $\gamma/\beta$ as a function of effective inverse temperature $\beta_{eff}$. The SYK model saturates the chaos bound $\gamma/\beta\leq 1/4$ at low temperature.}}
\label{gammabeta_plot}
\end{figure}
Figure \ref{gammabeta_plot} are the plots of $\gamma/\beta$ as a function of the effective inverse temperature $\beta_{eff}$. The SYK model saturates the chaos bound $\gamma/\beta\leq 1/4$ at low temperature. The XXZ spin chain becomes less chaotic at low temperature. This is expected because  low positive temperature means the left edge of the spectrum where it is known that the system becomes less chaotic in case of Hamiltonian with local interactions only \cite{Gubin_2012}.

Another motivation for considering the exponent $\gamma$ to be taken as a measure for chaos comes from the fact that smaller $\gamma$ leads to wider delocalization when the operator acts on an eigenstate. Cansider the system to be in an energy eigenstate $|E_n\rangle$. If we act on this wavefunction with the operator $A$, the new wavefunction would be a linear superposition of many energy eigenstates given by
\begin{equation}
A|E_n\rangle=\frac{1}{\mathcal{N}}\sum_m A_{mn}|E_m\rangle=\sum_m C_m|E_m\rangle\
\label{AEnstate}
\end{equation}
where $\mathcal{N}$ is the normalization factor. So, the new wavefunction is delocalized in the energy eigenbasis. This delocalization can be quantified in terms of inverse participation ratio (IPR) \cite{DAlessio:2015qtq} given by
\begin{equation}
IPR=\frac{1}{\sum_m|C_m|^4}
\end{equation}
IPR is large when the wavefunction is spread out in the particular eigenbasis under consideration which in our case is energy eigenbasis. As we have mentioned above, $f(\omega)$ is a constant for RMT. So, IPR will be maximum for the state in (\ref{AEnstate}) when we are working with RMT. RMT is the prototypical theory with which one compares a system to show that the system is chaotic. On the other hand, it is known that non-chaotic systems have small value of IPR for the above state because only a few $A_{mn}$ are large and non-zero. So, the delocalization of the above perturbed state quantified in terms of IPR can be considered as another measure of chaos. We have only quantified the degree of chaos further in terms of the exponent $\gamma$.

\begin{acknowledgments}
The author is fully supported by the Department of Science and Technology (Government of India) under the INSPIRE Faculty fellowship scheme\\(Ref. No. DST/INSPIRE/04/2020/002105).
\end{acknowledgments}

\bibliography{chaos_measure} 

\begin{thebibliography}{19}%
\makeatletter
\providecommand \@ifxundefined [1]{%
 \@ifx{#1\undefined}
}%
\providecommand \@ifnum [1]{%
 \ifnum #1\expandafter \@firstoftwo
 \else \expandafter \@secondoftwo
 \fi
}%
\providecommand \@ifx [1]{%
 \ifx #1\expandafter \@firstoftwo
 \else \expandafter \@secondoftwo
 \fi
}%
\providecommand \natexlab [1]{#1}%
\providecommand \enquote  [1]{``#1''}%
\providecommand \bibnamefont  [1]{#1}%
\providecommand \bibfnamefont [1]{#1}%
\providecommand \citenamefont [1]{#1}%
\providecommand \href@noop [0]{\@secondoftwo}%
\providecommand \href [0]{\begingroup \@sanitize@url \@href}%
\providecommand \@href[1]{\@@startlink{#1}\@@href}%
\providecommand \@@href[1]{\endgroup#1\@@endlink}%
\providecommand \@sanitize@url [0]{\catcode `\\12\catcode `\$12\catcode
  `\&12\catcode `\#12\catcode `\^12\catcode `\_12\catcode `\%12\relax}%
\providecommand \@@startlink[1]{}%
\providecommand \@@endlink[0]{}%
\providecommand \url  [0]{\begingroup\@sanitize@url \@url }%
\providecommand \@url [1]{\endgroup\@href {#1}{\urlprefix }}%
\providecommand \urlprefix  [0]{URL }%
\providecommand \Eprint [0]{\href }%
\providecommand \doibase [0]{https://doi.org/}%
\providecommand \selectlanguage [0]{\@gobble}%
\providecommand \bibinfo  [0]{\@secondoftwo}%
\providecommand \bibfield  [0]{\@secondoftwo}%
\providecommand \translation [1]{[#1]}%
\providecommand \BibitemOpen [0]{}%
\providecommand \bibitemStop [0]{}%
\providecommand \bibitemNoStop [0]{.\EOS\space}%
\providecommand \EOS [0]{\spacefactor3000\relax}%
\providecommand \BibitemShut  [1]{\csname bibitem#1\endcsname}%
\let\auto@bib@innerbib\@empty
\bibitem [{\citenamefont {Bohigas}\ \emph {et~al.}(1984)\citenamefont
  {Bohigas}, \citenamefont {Giannoni},\ and\ \citenamefont
  {Schmit}}]{Bohigas:1983er}%
  \BibitemOpen
  \bibfield  {author} {\bibinfo {author} {\bibfnamefont {O.}~\bibnamefont
  {Bohigas}}, \bibinfo {author} {\bibfnamefont {M.~J.}\ \bibnamefont
  {Giannoni}},\ and\ \bibinfo {author} {\bibfnamefont {C.}~\bibnamefont
  {Schmit}},\ }\href {https://doi.org/10.1103/PhysRevLett.52.1} {\bibfield
  {journal} {\bibinfo  {journal} {Phys. Rev. Lett.}\ }\textbf {\bibinfo
  {volume} {52}},\ \bibinfo {pages} {1} (\bibinfo {year} {1984})}\BibitemShut
  {NoStop}%
\bibitem [{\citenamefont {Maldacena}\ \emph {et~al.}(2016)\citenamefont
  {Maldacena}, \citenamefont {Shenker},\ and\ \citenamefont
  {Stanford}}]{Maldacena:2015waa}%
  \BibitemOpen
  \bibfield  {author} {\bibinfo {author} {\bibfnamefont {J.}~\bibnamefont
  {Maldacena}}, \bibinfo {author} {\bibfnamefont {S.~H.}\ \bibnamefont
  {Shenker}},\ and\ \bibinfo {author} {\bibfnamefont {D.}~\bibnamefont
  {Stanford}},\ }\href {https://doi.org/10.1007/JHEP08(2016)106} {\bibfield
  {journal} {\bibinfo  {journal} {JHEP}\ }\textbf {\bibinfo {volume} {08}},\
  \bibinfo {pages} {106}},\ \Eprint {https://arxiv.org/abs/1503.01409}
  {arXiv:1503.01409 [hep-th]} \BibitemShut {NoStop}%
\bibitem [{\citenamefont {Papadodimas}\ and\ \citenamefont
  {Raju}(2015)}]{Papadodimas_2015}%
  \BibitemOpen
  \bibfield  {author} {\bibinfo {author} {\bibfnamefont {K.}~\bibnamefont
  {Papadodimas}}\ and\ \bibinfo {author} {\bibfnamefont {S.}~\bibnamefont
  {Raju}},\ }\bibfield  {journal} {\bibinfo  {journal} {Physical Review
  Letters}\ }\textbf {\bibinfo {volume} {115}},\ \href
  {https://doi.org/10.1103/physrevlett.115.211601}
  {10.1103/physrevlett.115.211601} (\bibinfo {year} {2015})\BibitemShut
  {NoStop}%
\bibitem [{\citenamefont {Cotler}\ \emph {et~al.}(2017)\citenamefont {Cotler},
  \citenamefont {Gur-Ari}, \citenamefont {Hanada}, \citenamefont {Polchinski},
  \citenamefont {Saad}, \citenamefont {Shenker}, \citenamefont {Stanford},
  \citenamefont {Streicher},\ and\ \citenamefont {Tezuka}}]{Cotler_2017}%
  \BibitemOpen
  \bibfield  {author} {\bibinfo {author} {\bibfnamefont {J.~S.}\ \bibnamefont
  {Cotler}}, \bibinfo {author} {\bibfnamefont {G.}~\bibnamefont {Gur-Ari}},
  \bibinfo {author} {\bibfnamefont {M.}~\bibnamefont {Hanada}}, \bibinfo
  {author} {\bibfnamefont {J.}~\bibnamefont {Polchinski}}, \bibinfo {author}
  {\bibfnamefont {P.}~\bibnamefont {Saad}}, \bibinfo {author} {\bibfnamefont
  {S.~H.}\ \bibnamefont {Shenker}}, \bibinfo {author} {\bibfnamefont
  {D.}~\bibnamefont {Stanford}}, \bibinfo {author} {\bibfnamefont
  {A.}~\bibnamefont {Streicher}},\ and\ \bibinfo {author} {\bibfnamefont
  {M.}~\bibnamefont {Tezuka}},\ }\bibfield  {journal} {\bibinfo  {journal}
  {Journal of High Energy Physics}\ }\textbf {\bibinfo {volume} {2017}},\ \href
  {https://doi.org/10.1007/jhep05(2017)118} {10.1007/jhep05(2017)118} (\bibinfo
  {year} {2017})\BibitemShut {NoStop}%
\bibitem [{\citenamefont {Kudler-Flam}\ \emph {et~al.}(2020)\citenamefont
  {Kudler-Flam}, \citenamefont {Nie},\ and\ \citenamefont
  {Ryu}}]{Kudler_Flam_2020}%
  \BibitemOpen
  \bibfield  {author} {\bibinfo {author} {\bibfnamefont {J.}~\bibnamefont
  {Kudler-Flam}}, \bibinfo {author} {\bibfnamefont {L.}~\bibnamefont {Nie}},\
  and\ \bibinfo {author} {\bibfnamefont {S.}~\bibnamefont {Ryu}},\ }\bibfield
  {journal} {\bibinfo  {journal} {Journal of High Energy Physics}\ }\textbf
  {\bibinfo {volume} {2020}},\ \href {https://doi.org/10.1007/jhep01(2020)175}
  {10.1007/jhep01(2020)175} (\bibinfo {year} {2020})\BibitemShut {NoStop}%
\bibitem [{\citenamefont {Akutagawa}\ \emph {et~al.}(2020)\citenamefont
  {Akutagawa}, \citenamefont {Hashimoto}, \citenamefont {Sasaki},\ and\
  \citenamefont {Watanabe}}]{Akutagawa_2020}%
  \BibitemOpen
  \bibfield  {author} {\bibinfo {author} {\bibfnamefont {T.}~\bibnamefont
  {Akutagawa}}, \bibinfo {author} {\bibfnamefont {K.}~\bibnamefont
  {Hashimoto}}, \bibinfo {author} {\bibfnamefont {T.}~\bibnamefont {Sasaki}},\
  and\ \bibinfo {author} {\bibfnamefont {R.}~\bibnamefont {Watanabe}},\
  }\bibfield  {journal} {\bibinfo  {journal} {Journal of High Energy Physics}\
  }\textbf {\bibinfo {volume} {2020}},\ \href
  {https://doi.org/10.1007/jhep08(2020)013} {10.1007/jhep08(2020)013} (\bibinfo
  {year} {2020})\BibitemShut {NoStop}%
\bibitem [{\citenamefont {Xu}\ \emph {et~al.}(2020)\citenamefont {Xu},
  \citenamefont {Scaffidi},\ and\ \citenamefont {Cao}}]{Xu_2020}%
  \BibitemOpen
  \bibfield  {author} {\bibinfo {author} {\bibfnamefont {T.}~\bibnamefont
  {Xu}}, \bibinfo {author} {\bibfnamefont {T.}~\bibnamefont {Scaffidi}},\ and\
  \bibinfo {author} {\bibfnamefont {X.}~\bibnamefont {Cao}},\ }\bibfield
  {journal} {\bibinfo  {journal} {Physical Review Letters}\ }\textbf {\bibinfo
  {volume} {124}},\ \href {https://doi.org/10.1103/physrevlett.124.140602}
  {10.1103/physrevlett.124.140602} (\bibinfo {year} {2020})\BibitemShut
  {NoStop}%
\bibitem [{\citenamefont {Dowling}\ \emph {et~al.}(2023)\citenamefont
  {Dowling}, \citenamefont {Kos},\ and\ \citenamefont {Modi}}]{Dowling_2023}%
  \BibitemOpen
  \bibfield  {author} {\bibinfo {author} {\bibfnamefont {N.}~\bibnamefont
  {Dowling}}, \bibinfo {author} {\bibfnamefont {P.}~\bibnamefont {Kos}},\ and\
  \bibinfo {author} {\bibfnamefont {K.}~\bibnamefont {Modi}},\ }\bibfield
  {journal} {\bibinfo  {journal} {Physical Review Letters}\ }\textbf {\bibinfo
  {volume} {131}},\ \href {https://doi.org/10.1103/physrevlett.131.180403}
  {10.1103/physrevlett.131.180403} (\bibinfo {year} {2023})\BibitemShut
  {NoStop}%
\bibitem [{\citenamefont {Deutsch}(1991)}]{PhysRevA.43.2046}%
  \BibitemOpen
  \bibfield  {author} {\bibinfo {author} {\bibfnamefont {J.~M.}\ \bibnamefont
  {Deutsch}},\ }\href {https://doi.org/10.1103/PhysRevA.43.2046} {\bibfield
  {journal} {\bibinfo  {journal} {Phys. Rev. A}\ }\textbf {\bibinfo {volume}
  {43}},\ \bibinfo {pages} {2046} (\bibinfo {year} {1991})}\BibitemShut
  {NoStop}%
\bibitem [{\citenamefont {Srednicki}(1994)}]{Srednicki:1994mfb}%
  \BibitemOpen
  \bibfield  {author} {\bibinfo {author} {\bibfnamefont {M.}~\bibnamefont
  {Srednicki}},\ }\href {https://doi.org/10.1103/PhysRevE.50.888} {\bibfield
  {journal} {\bibinfo  {journal} {Phys. Rev. E}\ }\textbf {\bibinfo {volume}
  {50}},\ \bibinfo {pages} {888} (\bibinfo {year} {1994})},\ \Eprint
  {https://arxiv.org/abs/cond-mat/9403051} {arXiv:cond-mat/9403051}
  \BibitemShut {NoStop}%
\bibitem [{\citenamefont {Murthy}\ and\ \citenamefont
  {Srednicki}(2019)}]{Murthy:2019fgs}%
  \BibitemOpen
  \bibfield  {author} {\bibinfo {author} {\bibfnamefont {C.}~\bibnamefont
  {Murthy}}\ and\ \bibinfo {author} {\bibfnamefont {M.}~\bibnamefont
  {Srednicki}},\ }\href {https://doi.org/10.1103/PhysRevLett.123.230606}
  {\bibfield  {journal} {\bibinfo  {journal} {Phys. Rev. Lett.}\ }\textbf
  {\bibinfo {volume} {123}},\ \bibinfo {pages} {230606} (\bibinfo {year}
  {2019})},\ \Eprint {https://arxiv.org/abs/1906.10808} {arXiv:1906.10808
  [cond-mat.stat-mech]} \BibitemShut {NoStop}%
\bibitem [{\citenamefont {Sorokhaibam}(2024)}]{sorokhaibam2024quantum}%
  \BibitemOpen
  \bibfield  {author} {\bibinfo {author} {\bibfnamefont {N.}~\bibnamefont
  {Sorokhaibam}},\ }\href@noop {} {\bibinfo {title} {Quantum chaos and the
  arrow of time}} (\bibinfo {year} {2024}),\ \Eprint
  {https://arxiv.org/abs/2212.03914} {arXiv:2212.03914 [quant-ph]} \BibitemShut
  {NoStop}%
\bibitem [{\citenamefont {Foini}\ and\ \citenamefont
  {Kurchan}(2019)}]{Foini:2018sdb}%
  \BibitemOpen
  \bibfield  {author} {\bibinfo {author} {\bibfnamefont {L.}~\bibnamefont
  {Foini}}\ and\ \bibinfo {author} {\bibfnamefont {J.}~\bibnamefont
  {Kurchan}},\ }\href {https://doi.org/10.1103/PhysRevE.99.042139} {\bibfield
  {journal} {\bibinfo  {journal} {Phys. Rev. E}\ }\textbf {\bibinfo {volume}
  {99}},\ \bibinfo {pages} {042139} (\bibinfo {year} {2019})},\ \Eprint
  {https://arxiv.org/abs/1803.10658} {arXiv:1803.10658 [cond-mat.stat-mech]}
  \BibitemShut {NoStop}%
\bibitem [{\citenamefont {Sachdev}\ and\ \citenamefont
  {Ye}(1993)}]{Sachdev:1993}%
  \BibitemOpen
  \bibfield  {author} {\bibinfo {author} {\bibfnamefont {S.}~\bibnamefont
  {Sachdev}}\ and\ \bibinfo {author} {\bibfnamefont {J.}~\bibnamefont {Ye}},\
  }\href@noop {} {\bibfield  {journal} {\bibinfo  {journal} {Phys. Rev. Lett.}\
  }\textbf {\bibinfo {volume} {70}},\ \bibinfo {pages} {3339} (\bibinfo {year}
  {1993})}\BibitemShut {NoStop}%
\bibitem [{\citenamefont {Kitaev}(2015)}]{Kitaev:2015}%
  \BibitemOpen
  \bibfield  {author} {\bibinfo {author} {\bibfnamefont {A.~Y.}\ \bibnamefont
  {Kitaev}},\ }\href@noop {} {\bibinfo {title} {A simple model of quantum
  holography}} (\bibinfo {year} {2015}),\ \bibinfo {note} {talk at KITP
  Program: Entanglement in Strongly-Correlated Quantum Matter, University of
  California, Santa Barbara}\BibitemShut {NoStop}%
\bibitem [{\citenamefont {Maldacena}\ and\ \citenamefont
  {Stanford}(2016)}]{Maldacena:2016hyu}%
  \BibitemOpen
  \bibfield  {author} {\bibinfo {author} {\bibfnamefont {J.}~\bibnamefont
  {Maldacena}}\ and\ \bibinfo {author} {\bibfnamefont {D.}~\bibnamefont
  {Stanford}},\ }\href {https://doi.org/10.1103/PhysRevD.94.106002} {\bibfield
  {journal} {\bibinfo  {journal} {Phys. Rev. D}\ }\textbf {\bibinfo {volume}
  {94}},\ \bibinfo {pages} {106002} (\bibinfo {year} {2016})},\ \Eprint
  {https://arxiv.org/abs/1604.07818} {arXiv:1604.07818 [hep-th]} \BibitemShut
  {NoStop}%
\bibitem [{\citenamefont {Chowdhury}\ \emph {et~al.}(2022)\citenamefont
  {Chowdhury}, \citenamefont {Georges}, \citenamefont {Parcollet},\ and\
  \citenamefont {Sachdev}}]{Chowdhury_2022}%
  \BibitemOpen
  \bibfield  {author} {\bibinfo {author} {\bibfnamefont {D.}~\bibnamefont
  {Chowdhury}}, \bibinfo {author} {\bibfnamefont {A.}~\bibnamefont {Georges}},
  \bibinfo {author} {\bibfnamefont {O.}~\bibnamefont {Parcollet}},\ and\
  \bibinfo {author} {\bibfnamefont {S.}~\bibnamefont {Sachdev}},\ }\bibfield
  {journal} {\bibinfo  {journal} {Reviews of Modern Physics}\ }\textbf
  {\bibinfo {volume} {94}},\ \href
  {https://doi.org/10.1103/revmodphys.94.035004} {10.1103/revmodphys.94.035004}
  (\bibinfo {year} {2022})\BibitemShut {NoStop}%
\bibitem [{\citenamefont {Gubin}\ and\ \citenamefont
  {Santos}(2012)}]{Gubin_2012}%
  \BibitemOpen
  \bibfield  {author} {\bibinfo {author} {\bibfnamefont {A.}~\bibnamefont
  {Gubin}}\ and\ \bibinfo {author} {\bibfnamefont {L.~F.}\ \bibnamefont
  {Santos}},\ }\href {https://doi.org/10.1119/1.3671068} {\bibfield  {journal}
  {\bibinfo  {journal} {American Journal of Physics}\ }\textbf {\bibinfo
  {volume} {80}},\ \bibinfo {pages} {246} (\bibinfo {year} {2012})}\BibitemShut
  {NoStop}%
\bibitem [{\citenamefont {D'Alessio}\ \emph {et~al.}(2016)\citenamefont
  {D'Alessio}, \citenamefont {Kafri}, \citenamefont {Polkovnikov},\ and\
  \citenamefont {Rigol}}]{DAlessio:2015qtq}%
  \BibitemOpen
  \bibfield  {author} {\bibinfo {author} {\bibfnamefont {L.}~\bibnamefont
  {D'Alessio}}, \bibinfo {author} {\bibfnamefont {Y.}~\bibnamefont {Kafri}},
  \bibinfo {author} {\bibfnamefont {A.}~\bibnamefont {Polkovnikov}},\ and\
  \bibinfo {author} {\bibfnamefont {M.}~\bibnamefont {Rigol}},\ }\href
  {https://doi.org/10.1080/00018732.2016.1198134} {\bibfield  {journal}
  {\bibinfo  {journal} {Adv. Phys.}\ }\textbf {\bibinfo {volume} {65}},\
  \bibinfo {pages} {239} (\bibinfo {year} {2016})},\ \Eprint
  {https://arxiv.org/abs/1509.06411} {arXiv:1509.06411 [cond-mat.stat-mech]}
  \BibitemShut {NoStop}%
\end{thebibliography}%
\bibliographystyle{apsrev4-2}

\end{document}